\documentstyle[12pt]{article}
\begin{document}
\sloppy

\newcommand{\pa}{\partial}
\newcommand{\pab}{\bar{\partial}}
\newcommand{\zb}{\bar{z}}
\newcommand{\ZB}{\bar{Z}}
\newcommand{\dab}{\bar{D}}
\newcommand{\la}{\lambda}
\newcommand{\al}{\alpha}
\newcommand{\be}{\beta}
\newcommand{\th}{\theta}

\newcommand{\uz}{ {\bf z} }
\newcommand{\tb}{\bar{\theta}}
\newcommand{\thm}{\theta^-}
\newcommand{\tbm}{{\bar{\theta}}^-}

\newcommand{\bz}{{\bf Z}}

\newcommand{\re}{{\rm e}}
\newcommand{\ri}{{\rm i}}

\newcommand{\vF}{{\vec F}}

\newcommand{\half}{1\over 2}
\newcommand{\ph}{\phantom{-}}
\newcommand{\Tr}{{\rm Tr}\,}
\newcommand{\A}{{\cal A}}
\newcommand{\F}{{\cal F}}
\newcommand{\Z}{{\cal Z}}
\newcommand{\Zp}{{\cal Z}^{\prime}}

\newcommand{\beq}{\begin{equation}} 
\newcommand{\eeq}{\end{equation}} 
\newcommand{\bea}{\begin{eqnarray}} 
\newcommand{\eea}{\end{eqnarray}}
\newcommand{\de}{\delta}
\newcommand{\pr}{\prime}

\newcommand{\ds}{\displaystyle}
 
\newtheorem{theo}{Theorem}[section]
\newtheorem{lem}{Lemma}[section]
\newtheorem{prop}{Proposition}[section]
\newtheorem{cor}{Corollary}[section]
\newtheorem{defin}{Definition}[section] 
 
 
\hfill REF. TUW-00-05
\vskip 0.05truecm
\hfill LYCEN  2000-02
\vskip 0.07truecm
\hfill January 2000
 
\thispagestyle{empty}
 
\bigskip
\begin{center}
{\bf \LARGE{Conformal covariance}}
\end{center}
\begin{center}
{\bf \LARGE{in $2d$ conformal and integrable models,}}
\end{center}
\begin{center}
{\bf \LARGE{in $W$-algebras}}
\end{center}
\begin{center}
{\bf \LARGE{and in their supersymmetric 
extensions}\footnote{These notes, which are based in part  
on joint work with F.~Delduc,
S.~Gourmelen and S.~Theisen, 
represent the write-up of a talk given at the workshop 
 "Supersymmetries and Quantum
 Symmetries" (SQS'99, Dubna, July 27-31, 1999) organized by
 E.A. Ivanov, S.O. Krivonos and A.I. Pashnev.}}
\end{center}

\bigskip

\centerline{{\bf Fran\c cois Gieres}\footnote{On sabbatical leave from Institut 
de Physique Nucl\'eaire de Lyon,  
Universit\'e Claude Bernard, 43, boulevard du 11 novembre 1918, 
F-69622-Villeurbanne.}}

\bigskip
\centerline{{\it Institut f\"ur Theoretische Physik}}
\centerline{{\it Technische Universit\"at Wien}}
\centerline{{\it Wiedner Hauptstra\ss e 8-10}}
\centerline{{\it A-1040 Wien}}

\bigskip
\bigskip
\bigskip

\begin{abstract}
Conformal symmetry underlies 
the mathematical description of 
various two-dimensional  
integrable models (e.g. for their Lax representation, 
Poisson algebra, 
 zero curvature representation,...) or of conformal models 
 (for the anomalous Ward identities, operator product expansion,
Krichever-Novikov algebra,...)
   and of $W$-algebras. 
Here, we  review the construction of 
conformally covariant differential operators
which allow to render the conformal covariance 
manifest. The $N=1$ and $N=2$ 
   supersymmetric generalizations of these results 
 are also indicated and it is shown that they involve 
nonstandard matrix formats 
of  Lie superalgebras.

\end{abstract}
 
\newpage


\newpage

\noindent 
{\bf Remembering Viktor Ogievetsky}

\bigskip
 
Since the SQS'99 seminar is part of a series of International Seminars
dedicated to the 
 memory of  V.~Ogievetsky, I would like to recall briefly   
this great physicist and personality. 
I had the chance to meet 
Viktor when he came to CERN around 1989, and also later
on, when he was on visit at the University of Munich. 
The memories that I am keeping of these encounters are those 
of a very warm, kind and generous person, those of 
a physicist who was enthusiastic
about his work and always encouraging his colleagues, 
especially the younger ones, in their endeavor. 
Thus, I believe Viktor Ogievetsky 
should not only be remembered in science
through his important contributions to physics, 
but also through his general 
attitude towards research and all those 
involved in it.

\section{Introduction}

Conformal covariance is essential for the global formulation 
of scale invariant theories (conformal models)
on compact Riemann surfaces 
of any genus. 
It is at the heart of $W$-algebras 
which  are non-linear
generalizations of the two-dimensional conformal algebra, i.e. 
of the Virasoro algebra.
Moreover, as was realized in the eighties
and nineties, conformal symmetry 
manifests itself in several respects 
in two-dimensional integrable models 
like the KdV or Boussinesq equations.

By taking into account the underlying symmetries 
of a given theory, one  usually gains 
a better understanding of this 
theory \cite{dolo}. From a practical point of view, 
these symmetries generally provide a useful tool
for determining solutions or for checking results
within a given theory. 

Within the aforementioned theories and models, 
conformal symmetry 
manifests itself 
by the occurrence of conformally covariant 
differential operators in the time evolution equations 
or in the structure relations.  
In the present notes, we briefly 
review the definition and construction 
of these operators and of their supersymmetric extensions.

In our write-up, we have tried to maintain
the informal style of the oral presentation and therefore 
some results are only illustrated by the simplest 
examples. For more details,
we refer to the series of articles
\cite{cco1}-\cite{cco2} and to the work cited therein. 
(Among the latter, we explicitly mention 
references \cite{gp, dfiz, bfk} 
which represent the  basis for some 
parts of \cite{cco1}-\cite{cco2}.)
In reference \cite{cco2}, we 
illustrate how 
conformally covariant operators enter  
the physical models we mentioned and we show how   
they constrain, or largely determine, the form
of some of these theories.

\section{Geometric framework}

\subsection{Basic definitions} 

The arena we will work on, is a {\em Riemann surface} 
${\bf \Sigma}$,
i.e. a 
connected, topological $2$-manifold which is equipped 
with a {\em complex structure} (or equivalently, a real,
smooth,  connected and oriented $2$-manifold 
which is equipped 
with a {\em conformal class of metrics}) \cite{g}.
Roughly speaking, this means that 
any two systems of local complex coordinates, say $z$ and  
 $z^{\prime}$, are related by a conformal coordinate 
transformation, 
\[
z \ \stackrel{conf.}{\longrightarrow} \ z^{\prime} (z)
\ .
\]
In the following, we will use the notation 
$\pa \equiv {\pa \over \pa z}$
and we will denote the complex conjugate of $z$ by $\zb$.
Moreover, we assume that the considered Riemann surfaces 
are {\em compact} so that they are characterized by their 
{\em genus} $g\geq 0$.

A {\em conformal} (or {\em primary}) {\em field} 
of {\em weight} $k\in \bz /2$ 
on the Riemann surface ${\bf \Sigma}$ 
is a collection $\{ c (z , \zb) \}$ of local 
complex-valued functions on ${\bf \Sigma}$ 
(one for each coordinate system $(z, \zb )$),
transforming according to 
\beq
c^{\prime} (z^{\prime} , \zb ^{\prime} ) \, = \,  
(\pa z^{\prime})^{-k}
\, c(z , \zb)
\eeq
under a conformal change of coordinates. 
Thus, $c$ transforms linearly with a certain power of 
the Jacobian of the
change of coordinates\footnote{One can consider conformal
fields which also transform with a certain power of   
$\pab \zb ^{\prime}$, 
but we will not need them in the sequel.}. 
The space of conformal fields of weight $k$ on ${\bf \Sigma}$  
will be denoted by ${\cal F}_k$.  
 
The {\em Schwarzian derivative} 
of a conformal change of coordinates 
$z \to z^{\prime} (z)$ is defined by 
\beq
\label{sd}
S(z^{\prime} ; z) =
\pa^2 \, {\rm ln} \,
\pa z^{\prime}  - \frac{1}{2}
\left( \pa \, {\rm ln} \,
\pa z^{\prime}  \right) ^2
\ \ .
\eeq

A {\em projective} (or {\em Schwarzian})
{\em connection} \cite{g}
on the Riemann surface ${{\bf \Sigma}}$ is a
collection $\{ R (z , \zb ) \}$ of local 
complex-valued functions 
on ${\bf \Sigma}$ with the properties
 
(i) $R$ is
locally holomorphic,
i.e. $\pa_{\zb} R =0$,
 
(ii) $R$ transforms inhomogeneously with
the Schwarzian derivative
under a conformal change of coordinates
$z \rightarrow z^{\prime} (z)$ :
\begin{equation}
\label{601}
R^{\prime}  (z^{\prime}) \ = \  \left( \,
\pa z^{\prime}  \, \right) ^{-2}  \
\left[ \, R (z) \ - \ S (z^{\prime} ; z) \, \right]
\ \ .
\end{equation}
Such connections exist globally 
on compact Riemann surfaces of any genus.
From the physical point of view, the field $R$ and its complex 
conjugate represent the components of the energy-momentum tensor
in two-dimensional conformal field theory.

\subsection{Projective coordinates}
 
A change of local coordinates
$Z \rightarrow Z^{\prime} (Z )$ which has the
form
\begin{equation}
\label{333}
Z^{\prime} \ = \ \frac{aZ +b}{cZ+ d}
\ \ \ \ \ \ \ \ \
{\rm with} \ \ \ a,b,c,d \in {\bf C}
\ \ \ \ \
{\rm and} \ \ \ ad-bc =1
\ \ ,
\end{equation}
is called a {\em projective} (or {\em M\"obius} or {\em fractional
linear}) {\em transformation}. 
We note that the associated Jacobian is given by
\begin{equation}
\label{jac}
\pa_Z   Z^{\prime}  \ = \
(\, c Z + d\, )^{-2}
\ \ .
\end{equation}
In the following, coordinates belonging
to a projective atlas on the Riemann surface ${\bf \Sigma}$
will always be denoted by capital letters
$Z$ or $Z^{\prime}$. 

A {\em projective structure} on ${\bf \Sigma}$ is an atlas of local
coordinates for which all coordinate
transformations are projective. 
Every Riemann surface admits such a structure.
As a matter of fact, there is a 
{\em one-to-one correspondence
between projective structures and projective connections}
\cite{g}, see section 3.3 below.

Let ${\bf \Sigma}$ be a compact Riemann surface 
with a given projective structure. Then, 
a {\em quasi-primary field} of weight $k\in {\bf Z}/2$
on  ${\bf \Sigma}$ is a collection 
$\{ C_k ( Z ,\ZB )\}$ 
of local complex-valued functions on ${\bf \Sigma}$
which transform 
linearly with the $k$-th power of the Jacobian 
(\ref{jac}) under a projective change of coordinates:
 \begin{equation}
\label{cp}
C_k ^{\prime} \, =\,
( \, c Z +d  \, ) ^{2k}
\, C_k 
\ \ .
\end{equation}

\subsection{Covariant linear differential operators}

Consider the local form of a linear, holomorphic  
differential operator 
of order $n \in {\bf N}$, which is defined on the  Riemann 
surface ${\bf \Sigma}$:
\[
L^{(n)}=
a_0^{(n)} \pa^n +
a_1^{(n)} \pa^{n-1} +
a_2^{(n)} \pa^{n-2} + \dots +
a_n^{(n)} 
\qquad \ {\rm with} \ \ \ a_k^{(n)} = a_k^{(n)} (z)
\ .
\]
If the leading coefficient 
$a_0^{(n)}$ does not vanish anywhere, 
we can divide it. Therefore,
in the following,  we will 
assume that $a_0^{(n)} \equiv 1$.

\begin{defin}
A holomorphic, $n$-th order differential operator, which is locally 
 given on the compact Riemann surface ${\bf \Sigma}$ by 
\[
L^{(n)}=
\pa^n +
a_1^{(n)} \pa^{n-1} +
a_2^{(n)} \pa^{n-2} + \dots +
a_n^{(n)}
\ \ ,
\]
 is called {\em conformally covariant} if it maps conformal fields 
 (of some weight $p \in {\bf Z}/2$) to conformal fields: 
\[
L^{(n)} : {\cal F}_p \longrightarrow {\cal F}_{p+n}
\ \ .
\]
\end{defin}
This requirement is equivalent to the one that
$L^{(n)}$ transforms according to the following operatorial 
relation 
under a conformal 
change of coordinates 
$z \to z^{\prime} (z)$:
\beq
\label{trans}
L^{(n) \, \prime} =
(\pa z^{\prime})^{-(p+n)} \,
L^{(n)} \;
(\pa z^{\prime})^{p} \,
\ \ .
\eeq

According to the following result, the coefficient $a_1^{(n)}$
of a conformally covariant operator can always be eliminated 
without destroying conformal covariance \cite{cco2, bol}. 

\begin{theo}
Consider $n \in {\bf N}^{\ast}$.
 On a compact Riemann surface
 of genus 
$g >1$,  a conformally covariant operator $L^{(n)}$ for which 
the coefficient $a_1^{(n)}$ does not identically vanish, 
can only exist if it acts on conformal fields of weight 
\[
p = \frac{1-n}{2}
\ \ .
\]
In this case, $a_1^{(n)}$ transforms linearly under a conformal 
change of coordinates 
$z \to z^{\prime} (z)$, 
\beq
\label{a1}
a_1^{(n) \, \prime} =
(\pa z^{\prime})^{-1} \,
a_1^{(n)}
\ \ ,
\eeq
and thereby one can consistently  impose 
the vanishing of this coefficient.
The transformation law of $a_2^{(n)}$ then takes the simple form 
\beq
\label{a2} 
a_2^{(n) \, \prime} \; = \;
(\pa z^{\prime})^{-2} \left[
a_2^{(n)} - k_n
S(z^{\prime} ; z) \right]
\qquad \ where \ \ \ 
k_n \, =  \, \ds{n(n^2-1) \over 12} 
\ \ ,
\eeq
and where $S$ 
 denotes the Schwarzian derivative.   
\end{theo}

Accordingly, in the sequel, we will always consider 
conformally covariant operators which are normalized by 
$a_0^{(n)} \equiv 1, \, a_1^{(n)} \equiv 0$ and, for short, 
we will refer to these as CCO's:
\begin{defin} 
A \underline{CCO} (conformally covariant 
operator) of order $n$ on the compact Riemann surface 
${\bf \Sigma}$ is a map 
 \beq
\label{map}
L^{(n)} : {\cal F}_{ {1-n \over 2} } \longrightarrow 
 {\cal F}_{ {1+n \over 2} }
 \eeq
with the local expression 
 \beq
 \label{cco}
 L^{(n)}=
\pa^n +
a_2^{(n)} \pa^{n-2} + \dots +
a_n^{(n)}
\ \ . 
 \eeq
Here, the coefficients $a_2^{(n)},...,a_n^{(n)}$
are locally holomorphic functions on ${\bf \Sigma}$
and $a_2^{(n)}$ is a multiple of a projective connection:
\[
a_2^{(n)} \, = \, \ds{n(n^2-1) \over 12} \; R 
\ \ .
\]
\end{defin} 
The remaining coefficients $a_3^{(n)},...,a_n^{(n)}$
transform in a more complicated way than $R$ under conformal
changes of coordinates
\cite{dfiz}, so as to  
ensure the covariance (\ref{map}).

\section{CCO's}

From the conceptual point of view, CCO's are best approached  
by starting from 
the special coordinate system where $a_2^{(n)} =0$
 (i.e. by starting 
from projective coordinates $Z$) and then going over 
to generic local coordinates $z$ by a conformal transformation:
the dependence of the operators on the projective structure 
then translates into  
a  dependence on a projective connection. 
Therefore, we will first discuss operators on ${\bf \Sigma}$
which are covariant with respect to projective 
transformations.

\subsection{M\"obius covariant operators}

\medskip 

{\bf Class 1: Operators which only depend 
on the projective structure} 

\medskip

The operator 
$\pa_Z^n \equiv ({\pa \over \pa Z})^n$ 
(where $Z$ belongs to a projective atlas
on ${\bf \Sigma}$)
transforms homogeneously
if it acts on a quasi-primary field of weight ${1-n \over 2}$
\cite{gp}:
 
\begin{lem}[Bol's lemma]
Consider a projective
atlas on ${\bf \Sigma}$ with local changes of 
coordinates (\ref{333}).
If $C_{\frac{1-n}{2}} ( Z ,\ZB )$ is a quasi-primary
field on ${\bf \Sigma}$, then
$\pa_Z ^{n} \,
C_{\frac{1-n}{2}}$ also is, i.e. it transforms  
according to 
\begin{equation}
\left( \,
\pa_Z ^{n} \, C_{\frac{1-n}{2}} \, \right) ^{\prime}
\, = \,
\left( \, c Z +d  \, \right) ^{1+n} \
\pa_Z ^{n} \,  C_{\frac{1-n}{2}}
 \ \ \ \ \ ( \, n\, = \, 0, \, 1 , \, 2, \, ...\, )
\ \ .
\end{equation}
\end{lem}

\bigskip 

\noindent 
{\bf Class 2: Operators which depend linearly 
on a quasi-primary field}  

\medskip

For a given $n \in {\bf N}$ with $n\geq 3$, we consider linear 
M\"obius covariant operators $M^{(n)}_{W_3}, ...,M^{(n)}_{W_n} $ 
acting on 
quasi-primary fields of weight ${1-n \over 2}$.  
These operators do not only depend on the projective structure,
but also, in a linear way,  on quasi-primary fields 
$W_3,...,W_n$, respectively. Moreover, they are 
differential operators of lower order than $\pa^n_Z$.

Rather than giving a general formula for all of these operators 
(e.g. see \cite{cco1}), we present their explicit expression 
for $n=5$:
\begin{eqnarray}
\label{mw}
M^{(5)}_{W_5} & = & W_5
\qquad , \qquad
M^{(5)}_{W_{4}} \; = \; W_{4} \pa_Z + {1\over 2} (\pa_Z W_{4} )
\\
M^{(5)}_{W_{3}} & = & W_{3} \pa^2_Z 
+ (\pa_Z W_{3} ) \pa_Z
+ {2 \over 7} (\pa^2_Z  W_{3}) 
\ \ .
\nonumber
\end{eqnarray}

\subsection{From projective to generic coordinates}

Let us now go over 
from the projective coordinates $Z$ 
to generic holomorphic coordinates $z$ by a conformal transformation,
\[
Z \ \stackrel{conf.}{\longrightarrow} \ z
\ .
\]
In doing so, 
a quasi-primary field $C_k$ becomes a primary
field $c_k$, both fields being related by 
\begin{equation}
\label{ch}
C_k (Z, \ZB ) \, = \, (\pa  Z)^{-k} 
\, c_k (z, \zb ) 
\ \ .
\end{equation}
Moreover, a M\"obius covariant operator becomes a CCO. 
To discuss this passage, we  
consider in turn the two classes of examples introduced above.

\subsection{Class 1: Bol operators}

When passing from the projective coordinates $Z$ to the
holomorphic coordinates $z$ by a conformal transformation,
the $n$-th order derivative  
$\pa ^n _Z $ acting on a 
quasi-primary field $C_{1-n \over 2}$ 
becomes the $n$-th order {\em Bol operator} 
denoted by $L_n$:
\begin{equation}
\label{ch1}
\pa_Z ^n C_{\frac{1-n}{2}} \, = \,
(\pa Z) ^{-\frac{1+n}{2}} \;
L_n c_{\frac{1-n}{2}}
\ \ .
\end{equation}
By substituting the relation (\ref{ch}) with $k ={1-n \over 2}$
into equation (\ref{ch1}), we obtain the
following operatorial expression for the CCO $L_n$:  
\begin{equation}
\label{ch2}
L_n \ = \
(\pa Z) ^{\frac{1+n}{2}} \
\left( \frac{1}{\pa Z} \, \pa \right)^n \
(\pa Z) ^{-\frac{1-n}{2}}
\ \ .
\end{equation}
Thus, 
 the Bol operator   $L_n$ represents 
the  conformally covariant
version of the differential operator
$\pa^n$,  
the simplest examples
being given by
\begin{eqnarray}
L_0 & = & {\rm 1}
\nonumber  \\
L_1 & = & \pa
\nonumber  \\
L_2 & = & \pa^2 \ + \ \frac{1}{2} \, R
\nonumber  \\
L_3 & = & \pa^3 \ + \ 2 \, R \, \pa \ + \ (\pa R)
\label{81}
\\
L_4 & = & \pa^4 \ + \ 5 \, R \, \pa^2 \ + \ 5\, (\pa R)\, \pa \ + \
\frac{3}{2} \, \left[ (\pa^2 R) \, + \, \frac{3}{2} \, R^2  \right]
\ \ ,
\nonumber
\end{eqnarray}
where 
\begin{equation}
\label{a}
R_{zz} (z) \ \equiv \ S(Z ;z)
\ \ .
\end{equation}
This expression represents a projective connection, 
because it
has the correct transformation properties
thanks to the chain rule for the Schwarzian derivative.
From this chain rule, it also follows that
the definition (\ref{a}) is not affected by a
projective transformation of $Z$. Note that 
the quantity (\ref{a}) is holomorphic since 
the change of coordinates 
$z \rightarrow Z(z)$ has this property.
Equation (\ref{a}) expresses the
one-to-one correspondence
between projective structures and projective connections
that we already mentioned.

The basic operator $L_2$ (which is known as {\em Hill operator})
appears for instance in the Lax representation of the 
KdV equation while $L_3$ appears in the Poisson brackets
for the Virasoro algebra or in the conformal Ward identity
\cite{cco1}.

 \subsection{Class 2: Operators depending 
linearly on conformal fields}  

Upon passage   
$Z \to z$,
the quasi-primary field $W_k$ becomes a primary
field $w_k$, both fields being related by 
$W_k   =   (\pa  Z)^{-k} w_k$. Moreover, 
the M\"obius covariant operator  $M^{(n)}_{W_k}$
becomes a CCO
$M^{(n)}_{w_k}$ which depends  
linearly on $w_k$ and which acts on 
${\cal F}_{{1-n \over 2}}$. 
For instance, the $n=5$ operators 
(\ref{mw}) become
\begin{eqnarray}
\label{mwb}
M^{(5)}_{w_5} & = & w_5
\qquad , \qquad
M^{(5)}_{w_4} \; = \; w_4 \pa + {1\over 2} (\pa w_4  )
\\
M^{(5)}_{w_3 } & = & w_3 \left[ \pa^2 + 2  R \right]
+ (\pa w_3 ) \pa
+ {2 \over 7} \left[\pa^2  - 3 R \right]  w_3
\ \ .
\nonumber
\end{eqnarray}

 \subsection{Complete classification}  

 Any CCO 
\[
 L^{(n)}=
\pa^n +
a_2^{(n)} \pa^{n-2} + \dots +
a_n^{(n)}
\quad {\rm with }
\ \ a_2^{(n)} \,  =  \, \ds{n(n^2-1) \over 12}\;  R 
\]
can be 
reparametrized
in the following way in terms of the projective connection $R$ 
and $n-2$ conformal fields $w_3,...,w_n$:
\beq
\label{cp1}
L^{(n)} = L_n + M^{(n)}_{w_3} + ... + M^{(n)}_{w_n}
 \ \ .
\eeq
The relation between the coefficients $a_3^{(n)},...,a_n^{(n)}$ 
and the conformal fields $w_3,...,w_n$ is given by differential 
polynomials which involve $R$ and this relation 
is invertible. 

The parametrization (\ref{cp1}) of $L^{(n)}$ in terms of the 
energy-momentum tensor and some conformal fields is very
helpful for the construction and formulation of $W_n$-algebras,
see section 3.7 below.

\subsection{Nonlinear conformally covariant operators}

There exists a unique bilinear conformally covariant 
operator $J(\cdot,\cdot)$, the so-called {\em Gordan transvectant}
\cite{gp, dfiz, cco1}. Here, we only note that it  
encompasses the CCO's $M^{(n)}_{w_k}$: 
\begin{equation}
M^{(n)} _{w_k} c \, \propto \, J ( w_k, c) 
\ \ .
\end{equation}
The bilinear operator $J(\cdot,\cdot)$ as well as  
higher multilinear conformally covariant  
operators appear in the  
defining  relations of $W$-algebras \cite{gt,cco2}.

\subsection{Matrix representation of CCO's}

The CCO's $L_n$ and
$M^{(n)}_{w_k}$ 
admit a matrix representation which is related to the
principal embedding of the Lie algebra 
$sl(2)$ into $sl(n)$ \cite{dfiz}. 
Since  
$sl(2)$ is the Lie algebra of the M\"obius group, this 
algebraic relationship which underlies the matrix representation 
of CCO's,
reflects the fact that these covariant operators 
come  from 
M\"obius covariant ones. We will now illustrate the matrix 
representation for $L^{(3)} = L_3 + M^{(3)}_{w_3}= 
L_3 + w_3$. 

Let us rewrite the scalar, conformally covariant 
differential equation
\beq
\label{start}
0= L^{(3)} f_3 
\equiv \left[
\pa ^3 + 2 R  \pa + (\pa R) + w_3 \right] f_3
\qquad  {\rm with}
\ f_3 \in {\cal F}_{-{1\over 2}} 
\eeq
as a system of
three first-order differential equations:
\beq 
\label{f}
\left[
\begin{array}{c}
0 \\ 0 \\ 0
  \end{array}
   \right]
= 
\left[
\begin{array}{ccccc}
\pa  & R    & w_3  \\
-1   & \pa &   R   \\
 0   & -1  &   \pa   
   \end{array}
   \right]
\left[
\begin{array}{c}
f_1 \\ f_2 \\ f_3
  \end{array}
   \right]
\quad \Longleftrightarrow \quad \left\{
\begin{array}{l}
0= \pa f_1 + Rf_2 + w_3 f_3 
\\
f_1 = \pa f_2 + Rf_3 
\\
f_2 = \pa f_3 \ .
  \end{array}
\right.
\eeq
Substitution of the last two equations into the first one
reproduces the scalar equation 
(\ref{start}). 

Equation (\ref{f}) can also be written in the form 
\beq
\vec 0 = ( \pa - {\cal A}) \vec F
\qquad {\rm with } \ \ 
{\cal A} =
 \left[
\begin{array}{ccccc}
0 & -R    & -w_3  \\
1   & 0 &   -R   \\
 0   & 1  &   0   
   \end{array}
   \right]
\ \ , \ \ 
\vec F = 
\left[
\begin{array}{c}
f_1 \\ f_2 \\ f_3
  \end{array}
   \right]
\ \ .
\eeq
Here, the matrix ${\cal A}$ can be viewed as the $z$-component 
of a two-dimensional 
gauge connection with values in the Lie algebra $sl(3)$.
After supplementing ${\cal A}$ with a $\zb$-component, 
one can derive the $W_3$-algebra
by imposing a 
{\em zero curvature condition} on the connection
\cite{bfk,cco2}.

\section{$N=1$ supersymmetry}

\subsection{General framework}

The $N=1$ supersymmetric generalization of the previous results
has
been worked out in references \cite{cco1, gt} 
(see also \cite{hua}) by using  a superspace approach.
We note that 
$N=1$ superspace is locally parametrized by complex 
coordinates $z$ and $\th$ which are even and odd, respectively.
The transition from 
ordinary space to superspace  
can be summarized as follows:
\begin{equation}
\left.
\begin{array}{l}
\mbox{Riemann surface} \\ 
z \\
\pa  \\ 
\mbox{conformal transformation} \\  
\mbox{conformal field} \ c_k \\  
\mbox{projective connection} \ R_{zz}(z) \\  
 L_2 = \pa^2 + {1 \over 2} R_{zz}  
\end{array}
\right\} 
\ \rightarrow \ \left\{
\begin{array}{l}
\mbox{super Riemann surface} \\ 
z, \th \\
\pa , D \equiv {\pa \over \pa \th } + \th \pa \quad (D^2 =\pa ) \\ 
\mbox{superconformal transf.} : D z ^{\prime} = 
\th ^{\prime} D \th ^{\prime}       \\   
\mbox{superconformal field}:
\  {\cal C}_k ^{\prime} = 
(D\th^{\prime})^{-k} {\cal C}_k  \\  
\mbox{superprojective connection} \ {\cal  R} _{z\th} (z,\th)  \\ 
{\cal  L}_1 = D^3+ {\cal  R}_{z\th}
\ \ . 
\end{array}
\right.
\end{equation}
The odd superdifferential operator 
${\cal  L}_1$ acts on a superconformal field 
${\cal C}_{-1} \equiv {\cal C}$.  
By applying $D$ to ${\cal  L}_1 {\cal C}$ 
and subsequently projecting onto the lowest 
component of the resulting superfield, we find
\begin{eqnarray*}
(D {\cal  L}_1 {\cal C})| &=& 
\left[ D^4 + (D {\cal  R}_{z\th})| \right] {\cal C}|
- {\cal  R}_{z\th}|  (D {\cal C}_3)| \\
 &=& [ \pa^2 + {1 \over 2} R_{zz}  ] c +
\rho_{z\th}  (D {\cal C})|
\ \ ,
\end{eqnarray*}
i.e. the basic Bol operator $L_2$ plus a fermionic
contribution. 

\subsection{Matrix representation of super CCO's}

Let us rewrite the scalar, superconformally covariant 
differential equation
\beq
\label{sst}
0= {\cal L}_1 F_3 
\equiv \left[
D^3 +  {\cal R}  \right] F_3
\eeq
as a system of
three first-order differential equations:
\beq 
\label{sus}
\left[
\begin{array}{c}
0 \\ 0 \\ 0
  \end{array}
   \right]
= 
\left[
\begin{array}{ccccc}
D  & 0   & {\cal R} \\
-1   & D &   0   \\
 0   & -1  &   D  
   \end{array}
   \right]
\left[
\begin{array}{c}
F_1 \\ F_2 \\ F_3
  \end{array}
   \right]
\quad \Longleftrightarrow \quad \left\{
\begin{array}{l}
0= D F_1 + {\cal R} F_3 
\\
F_1 = D F_2  
\\
F_2 = D F_3 \ .
  \end{array}
\right.
\eeq
Analogously to the non supersymmetric theory, 
substitution of the last two equations into the first one
reproduces the scalar equation 
(\ref{sst}). 

If we now rewrite equation (\ref{sus}) in the form 
$\vec 0 = (D - {\cal A}) \vec F$, we realize that the 
matrix ${\cal A}$ belongs to the Lie superalgebra 
$sl(2|1)$. However, the graded matrix ${\cal A}$
does not have the standard format which consists of 
arranging 
the even and odd matrix elements  into blocks:
this is an example of a {\em nonstandard matrix format}, to which 
we have referred as the {\em diagonal format}  since 
there are alternatively 
even and odd diagonals \cite{gt,dg}. 

This and  other possible 
nonstandard matrix formats have been studied in a systematic
way in reference \cite{dg}. Although they are simply related 
to the standard format by a similarity transformation, 
they have many appealing features. 
Moreover, such formats naturally occur
in various physical applications, 
e.g. in superconformal field theory, 
superintegrable models, for super $W$-algebras 
and quantum supergroups.

\section{$N=2$ supersymmetry}

A $N=2$ super Riemann surface is locally parametrized 
by an even complex coordinate $z$ and two  
odd complex coordinates $\th$ and $\tb$. 
There is a new feature in $N=2$ superspace geometry 
which makes this
theory considerably richer and more complicated than 
the $N=1$ supersymmetric theory: 
the ``square root" of the translation generator $\pa$
is not given by a single odd operator as in $N=1$ supersymmetry
($D^2 = \pa$), but it involves two odd operators, 
\beq
\label{1}
D  \, =  \, \frac{\pa}{\pa \th}  +  \frac{1}{2} \, \tb \pa
\qquad \ , \qquad  
\dab \,  = \,  \frac{\pa}{\pa \tb}  +  \frac{1}{2} \, \th  \pa
\ \ ,
\eeq
satisfying 
\begin{equation}
\label{2}
\{  D ,  \dab \} =  \pa 
\end{equation}
(and 
$D^2   =  0  = \dab^2$).
Therefore, one has to deal with partial differential equations
(involving $D$ and $\dab$) rather than ordinary differential
equations (only involving $D$).
Another aspect of the algebra
$ \{ D , \dab \} = \pa$ consists of the fact that it introduces a
$U(1)$ symmetry into the theory: after projection from the
super Riemann surface to the underlying ordinary Riemann
surface, one thereby recovers $U(1)$-transformations
in addition to the familiar conformal transformations. Henceforth,
the Bol operators (\ref{81}) acting on $U(1)$-neutral fields
are to be generalized to conformally covariant operators
acting on $U(1)$-charged fields.
The latter as well as the original operators
(\ref{81}) arise from different types of 
$N=2$ super Bol operators which have been  
constructed and classified in reference \cite{gs}.
For a particular class of them, 
the so-called `sandwich operators' 
(relating the chiral and anti-chiral subspaces
of superconformal fields), 
one can give a matrix representation. 
The results following from a zero curvature condition 
for the operator product 
expansions of the $N=2$ super $W_3$-algebra 
coincide with those obtained by other methods
\cite{iva}.

\section{Concluding comments}

In these notes, we have tried to give a short introduction 
to some mathematical notions which 
are quite useful for the study of many physically interesting 
models in two dimensions. 
While the appearance of conformal symmetry in conformal 
models or in their non-linear  generalizations
(related to $W$-algebras) is quite natural, the role 
of conformal invariance in integrable models is less clear 
and still a matter of current research 
(see references \cite{hi} and the contribution of M.Olshanetsky
to this workshop).

\newpage 
 
{\bf \Large Acknowledgments}
 
\vspace{3mm}
 
I wish to thank my collaborators with whom I have worked 
on the topics reviewed in this talk, 
for our pleasant collaborations. 
Furthermore,
I express my gratitude to Evgeny Ivanov, Sergey Krivonos 
and Anatoly Pashnev for their kind invitation to participate  
in the workshop,
for nicely taking care of all practical matters and 
 for rendering my visit to Dubna a most pleasant 
and rewarding one. The final version of this text was prepared 
while I was on sabbatical leave at the 
Technical University of Vienna: 
I wish to thank
Prof.M.Schweda for his kind invitation and all the members 
of the Institut f\"ur Theoretische Physik 
for the hospitality extended to me.

\end{document}